\documentclass[
 reprint,
 amsmath,amssymb,
 aps,
]{revtex4-1}

\usepackage{graphicx}
\usepackage{dcolumn}
\usepackage{bm}

\begin{document}

\preprint{APS/123-QED}

\title{The instability of followers and emergent vorticity in\\ flocking behaviour for an experimental interaction rule}

\author{Michael Small}
\affiliation{School of Mathematics and Statistics\\The University of Western Australia, Crawley, WA, Australia, 6009}
\email{michael.small@uwa.edu.au}
\author{Xiao-ke Xu}%
 \email{xiaokeeie@gmail.com}
\affiliation{College of Information and Communication Engineering, Dalian Nationalities University, Dalian, 116605, China}

\date{\today}

\begin{abstract}
Computational models of collective behavior in birds has allowed us to infer interaction rules directly from experimental data. Using a generic form of these rules we explore the collective behavior and emergent dynamics of a simulated swarm. For a wide range of flock size and interaction extent (the fixed number of neighbors with which an individual will interact) we find that the computational collective is inherently stable --- individuals are attracted to one another and will position themselves a preferred distance from their fixed neighbors within a rigid lattice. Nonetheless, the irregular overall shape of the flock, coupled with the need for individuals on the boundary to move towards their neighbors creates a torque which leads the flock to rotate and then meander. We argue that this ``rolling meander'' is a very good proxy for real collective behavior in animal species and yet arises from a simple homogeneous and deterministic rule for interaction. Rather than then introduce leaders --- which has already been shown, quite straightforwardly, to drive collective swarms such as this --- we introduce a small number of ``followers''. Each follower is bound to consider a random fixed individual to be among their neighbors, irrespective of actual metric distance between them. We find that the introduction of a small number of such followers causes a phase transition that quickly leads to instability in the flock structure (as no stable configuration arises) and the previously rigid crystalline interaction among neighbors now becomes fluid: the distance between neighbors decreases, the flock ceases to rotate and meanders less.  \end{abstract}

\pacs{Valid PACS appear here}
\maketitle

Collective behavior occurs widely in nature, and as a consequence efforts to understand it have been vigorous \cite{dS10}. Physicists and computer scientists have had their own unique interest in this problem as simple mathematical models for inter-particle interaction have been shown to lead to interesting collective behaviors \cite{cR87} and even phase transitions as a function of particle density \cite{tV95}. While biologists approach this problem with efforts to build intricate models of inter-individual interactions that capture the known biology, physicists have focussed on these {\em ad hoc} models with simple mechanics that produce ``naturalistic'' behaviors. Recent advances in tracking and monitoring technology now allows for a third approach. By observing real flocks it is possible to build models of inter-individual interaction directly from the data \cite{mN10}. In \cite{chano1} we did this for data from a group of homing pigeons in free flight. One of the observations of this work was that the hierarchical leadership relationship detected originally \cite{mN10} was actually supplemented by a more egalitarian reciprocity \cite{xu5}. 

In this communication we take these models built from pigeon flight data and ask what features can we observe in the models which are necessary to explain observed naturalistic flight? And, what dynamical phenomena do these models exhibit? This is a different problem from that treated by Vicsek \cite{tV95} and Reynolds \cite{cR87} --- for whom the model was created with a combination of intuition and foresight. It is also different from what has been done in biology \cite{mN10,aW11,mB08}, as we do not ground our model directly on theory but rather on experimental observation and data. Our work diverges still further from the approach taken in biology as we distill from the computational model a simple interaction rule consistent with both the observed interaction data and collective behavior.

\begin{figure}
\begin{center}
\begin{tabular}{cc}
\includegraphics[width=0.225\textwidth]{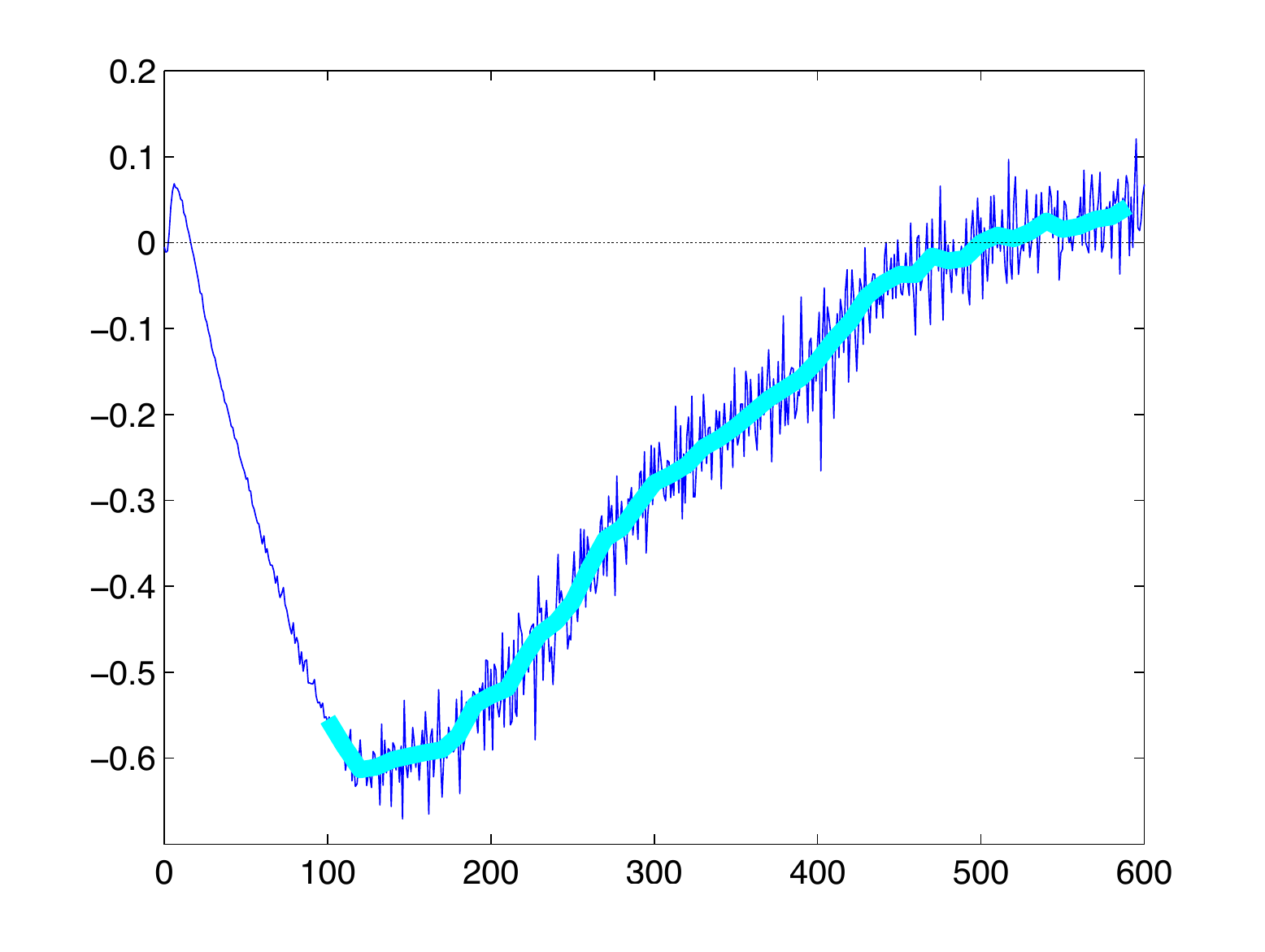}
&\includegraphics[width=0.18\textwidth]{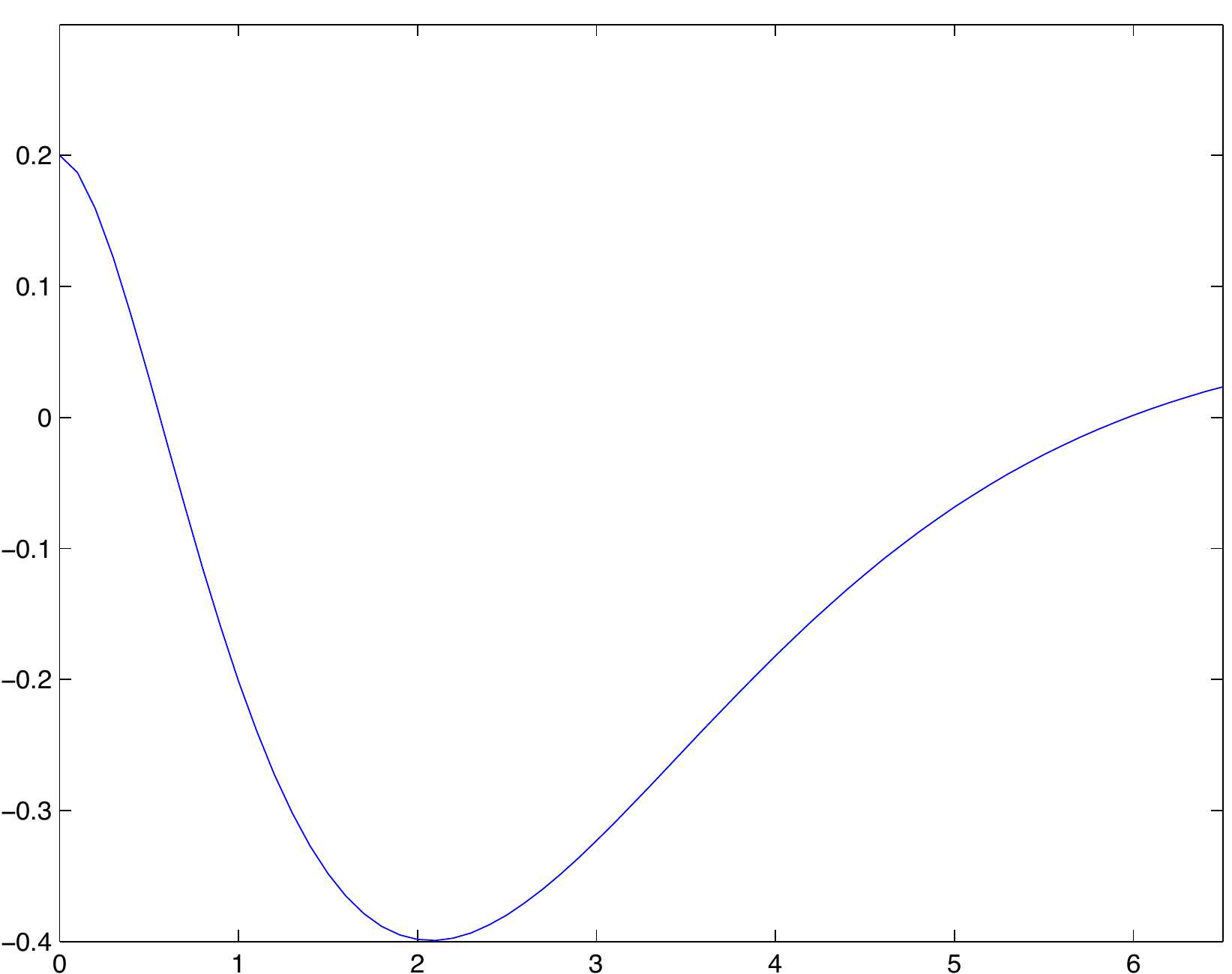}\\
(a) &(b)
\end{tabular}
\end{center}
\caption{{\bf Pairwise attraction repulsion.} [Colour Online]  Panel (a) depicts the pairwise force of attraction or repulsion between two particles based on simulations of the computational model built from the data --- following methods described extensively in \cite{chano1}. Panel (b) is the simplified proxy for this curve $h(x)=\frac1{x+a}-x^be^{-x}$ (for $a=5$ and $b=2$) which we will employ in the current work. }
\label{hforce}
\end{figure}

The details of our modeling procedure have been described elsewhere \cite{chano1} and, while foundational, are not directly relevant to the current discussion. Nonetheless, we build a model of particle-particle interaction where the position $x_t^{(i)}$  of particle $i$ is updated as a function  $f(\cdot)$ of its current position and velocity and those of its $m$ nearest neighbors. The function $f$ is chosen, via a numerical procedure \cite{kJ95a,model_thing,mdlnn}, to fit the observed data. The exact details of this procedure are not relevant here and any of the many alternatives would do just as well. Nonetheless, from this model $f$ we can now estimate the strength of pairwise interaction between two particles. If the distance between two particles $i$ and $j$ is given by $d_{ij}=\|x^{(i)}-x^{(j)}\|$ then we define the force of attraction $\tilde h(d_{ij})$. The form of $\tilde h(\cdot)$ can be estimated from the data and is shown in Fig. \ref{hforce}.(a).

The exact (that is, exact representation of the model $f$) interaction rule $\tilde h$ is computationally complicated. In Fig. \ref{hforce}(b) we depict the proxy we will use here instead: $h$. In addition to being simpler to compute, $h$ has the advantage that it is more easily replicable. While we could use the exact curve depicted in Fig. \ref{hforce}(a), this is not actually necessary. It is only the general shape which we wish to capture and hence we use the function
\begin{eqnarray}
h(x) & = &\frac 1{x+a}-x^be^{-x}
\label{heq}
\end{eqnarray}
as a computationally advantageous expediency. With $h(x)$ providing an attractive/repuslve force for the interaction between two particles based on distance, we are now able to state the swarm dynamic which we study here.

Let the swarm consist of $N$ particles such that $x_t^{(i)}$ is the position and $v_t^{(i)}$ the velocity of particle $i$ at time $t$. Each particle updates its velocity and position based on interaction with its $k$ nearest neighbors. Denote by $x_t^{i\leftarrow j}$ the $j-th$ nearest neighbor of particle $i$ at time $t$. We then compute two competing {\em desired}  velocity vectors for each particle \begin{eqnarray}
\label{vh} \hat{v}_H^{(i)} & = & \frac{1}{k}\sum_{j=1}^kh(d_{ij})(x_t^{i\leftarrow j}-x_t^{(i)})\\
\label{vv} \hat{v}_V^{(i)}  & =& \frac{1}{k+1}\left(v_t^{(i)}+ \sum_{j=1}^kv_t^{i\leftarrow j}\right)\\
\nonumber d_{ij} &:=& \|x_t^{i\leftarrow j}-x_t^{(i)}\|
\end{eqnarray}
where, for each particle, $\hat{v}_V$ is the desired velocity from velocity alignment amount the neighbors and $\hat{v}_H$ is the desired velocity based on the pairwise attraction or repulsion with each neighbor. The actual updated velocity is a convex combination of these two desired quantities, and the position is then computed from this:
\begin{eqnarray}
v_{t+1}^{(i)} & = & \lambda\hat{v}_V^{(i)} +(1-\lambda)\hat{v}_{H}^{(i)}\\
x_{t+1}^{(i)} & = & x_t^{(i)}+\epsilon{v}_{t+1}^{(i)}
\end{eqnarray}
where $0\leq\lambda\leq 1$ and $\epsilon$ is an appropriate integration time step. In what follows we have set $\lambda=0.3$, but this choice is not critical, extensive repeated simulations with other intermediate (i.e. $\lambda\neq 0,1$) values provide equivalent results.

The qualitative behaviors of simulated swarms following these rules are striking, and have been briefly discussed in the abstract. For all values of $N$ (up to $5000$) and $k$ ($1\leq k\leq 28$) our simulations indicate that the swarm will find a stable configuration --- initially randomly distributed particles (both in space and velocity) quickly rearrange themselves into a stable configuration reminiscent (for smaller $k$) of a  sub-optimal sphere packing (each particle strives to reach its preferred distance from the requisite neighbors). The position of each particle relative to its neighbors then remains fixed within  a single cloud. However, since particle on the boundary of that cloud have neighbor to only one side of them, these particles attempt to move towards their neighbors in such a way that the irregular shape of the boundary creates a non-zero rotational torque. The cloud rotates and then meanders.

\begin{figure}
\begin{center}
\begin{tabular}{cc}
\includegraphics[width=0.45\textwidth]{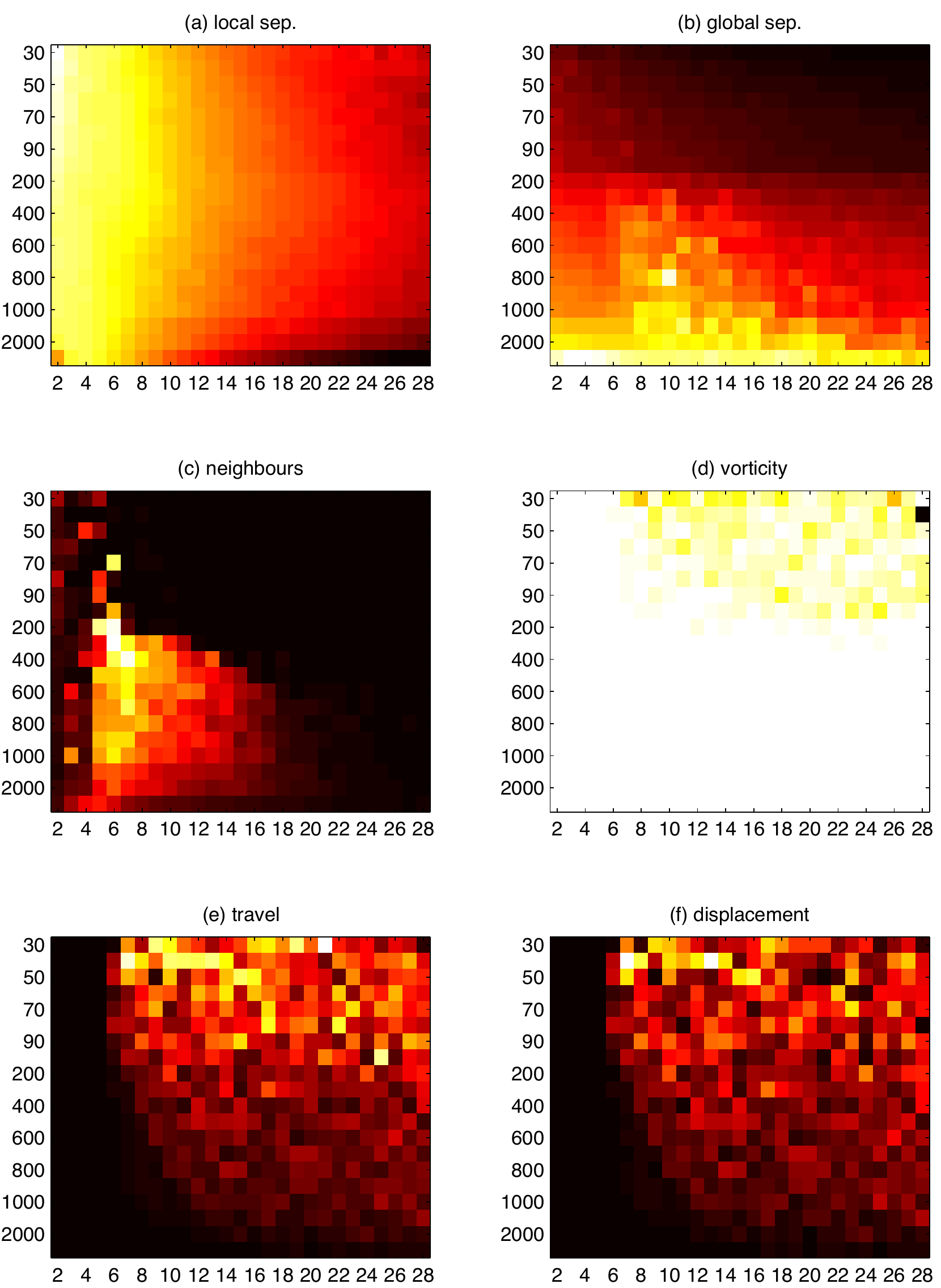}
\end{tabular}
\end{center}
\caption{{\bf Effect of swarm size and number of neighbors} [Colour Online]  Panel (a-f) depicts heat maps and/or contour plots for each of the size parameters defined in the text (local separation, global separation, neighbor stability, vorticity, travel and movement) for $k=1,2,\ldots 28$ and $100\leq N\leq 5000$. Darker coloring corresponds to smaller numerical values.}
\label{gbeh}
\end{figure} 

Fascinating as this rigid form of self organization is, we prefer to pursue   a more quantitative analysis here. To do so we define the following six quantities:
\begin{itemize}
\item {\em local separation: } the average over all particles in the flock of the distance between that point and its $k$ nearest neighbors;
\item{\em global separation:} the maximum distance between any two particles in the swarm;
\item{\em neighbor stability:} the average over all particles of number of different particles which become the nearest neighbor of a given particle over a fixed time interval;
\item{\em vorticity:} the mean rotation, between successive time intervals of each particle around the centre of mass of the swarm;
\item{\em travel:} the path integral of the centre of mass over a time interval; and,
\item{\em movement:} the straight line distance travelled over the same time period.
\end{itemize}
For a wide range of values of $k$ and $N$ we simulated the behaviour of swarms of particles from random initial conditions --- random position and velocity. In each case we simulated for $20000$ time steps and computed the parameters defined above over the last 500 (sampling only everything 10-th time step) configurations. In Fig. \ref{gbeh} we depict the behaviour of each of these six parameters.

\begin{figure}
\begin{center}
\begin{tabular}{cc}
\includegraphics[width=0.45\textwidth]{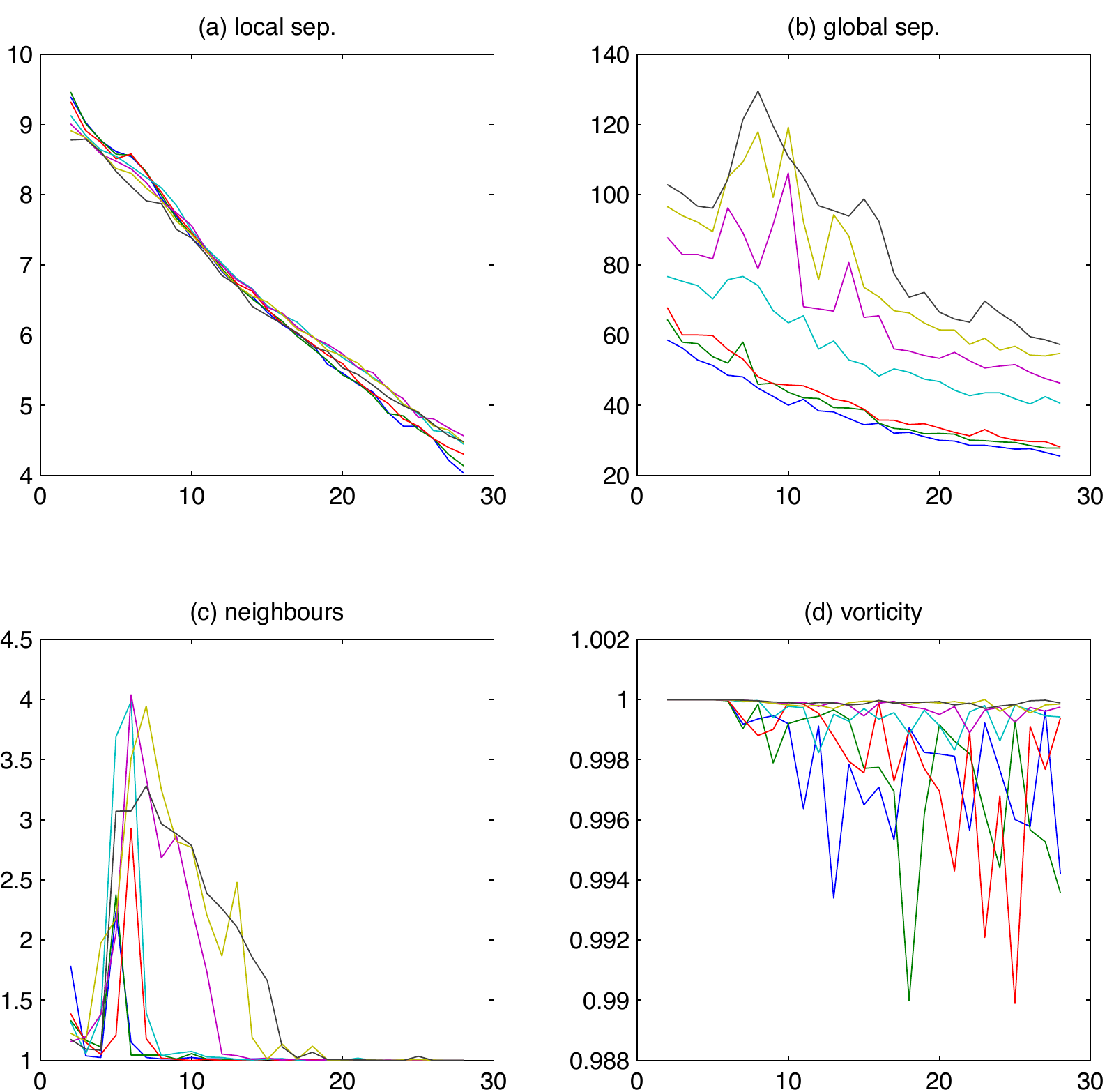}
\end{tabular}
\end{center}
\caption{{\bf Phase transition} [Colour Online]  (a) local separation, (b) global separation, (c) neighbor stability, and (d) vorticity as a function of $k=1,2,\ldots 28$ for $N=80, 90, 100, 200, 300, 400, 500$. Behaviour of local and global separation is as describe above for Fig. \ref{gbeh}. For neighbor stability there is a clear change in stability as $k$ increases the swarm ceases to have a rigid equilibrium and the nearest neighbors constantly switch, for larger still $k$($k>20$) the stable rigid structure returns. The width of the constant window increases with $N$ (over the range of $N$ plotted here): $N=300, 400, 500$ show the largest deviation from a stable rigid structure. Vorticity shows a clear increase are $k$ exceeds about 6 --- beyond this point the swarm will start to rotate and continue to do so. Significant variability between simulations is particularly evident here.  }
\label{phase}
\end{figure}

Unsurprisingly, local separation depends primarily on the number $k$ (imagine a single ball trying to be equally close to $k$ other balls as $k$ increases) and global separation depends on $N$. Nonetheless, we do see for a fixed $k$ a slight variance in local separation as function of $N$ --- separation is greatest around $N=400$. Similarly, for a fixed $N$ global separation is decreased by increasing $k$: as a larger $k$ introduces a stronger cohesive force: nonetheless, there is a slight peak in global separation for intermediate values of $k$ (around $k=10$, see Fig. \ref{phase} (b)). The plot of neighbor stability is complementary to these findings as we see that for small $k$, and more evidently for larger $N$ we observe a range of values of $k$ for which a stable packing cannot be found (a stable configuration corresponds to a small value of neighbor stability and hence a darker coloration in Fig. \ref{gbeh}(c)). Nonetheless, as $k$ increases, more stable configurations dominate. Conversely, greater vorticity (indicated by a smaller numerical value, and hence darker coloration, as this number is the cosine of the angular change) is evident for smaller $N$ and larger $k$ --- and coincides with the values of neighbor stability which are stable. That is, a stable configuration will start to rotate, whereas an unstable one does not. Both travel and movement (as defined above) indicate similar results --- stable configurations rotate and hence move more. There is no pattern in the ratio of travel to movement (not shown) as both quantities provide very similar information.

In Fig. \ref{phase} we depict the phase transition more clearly by illustrating the value of these quantities as a function of only one of $k$ or $N$. This figure displays the same data as in Fig. \ref{gbeh}, when viewed together it is clear that one of two distinct behaviors persists in this system. Either a rigid fixed near neighbor structure with a rotating swarm and overall meander or, a swarm in which individuals are constantly vying for position and lacking rotation or general drift.  That is, we observe a transition from a rigid crystalline configuration (with each particle fixed in its relative position to its neighbors) to a gaseous phase (with particle constantly rearranging and moving past one another. The rigid configuration also exhibits a rotational force and hence a general meander. Nonetheless, the gaseous phase is only transitional, by increasing $k$ further (for fixed $N$) the crystalline configuration resumes.

\begin{figure}
\begin{center}
\begin{tabular}{cc}
\includegraphics[width=0.45\textwidth]{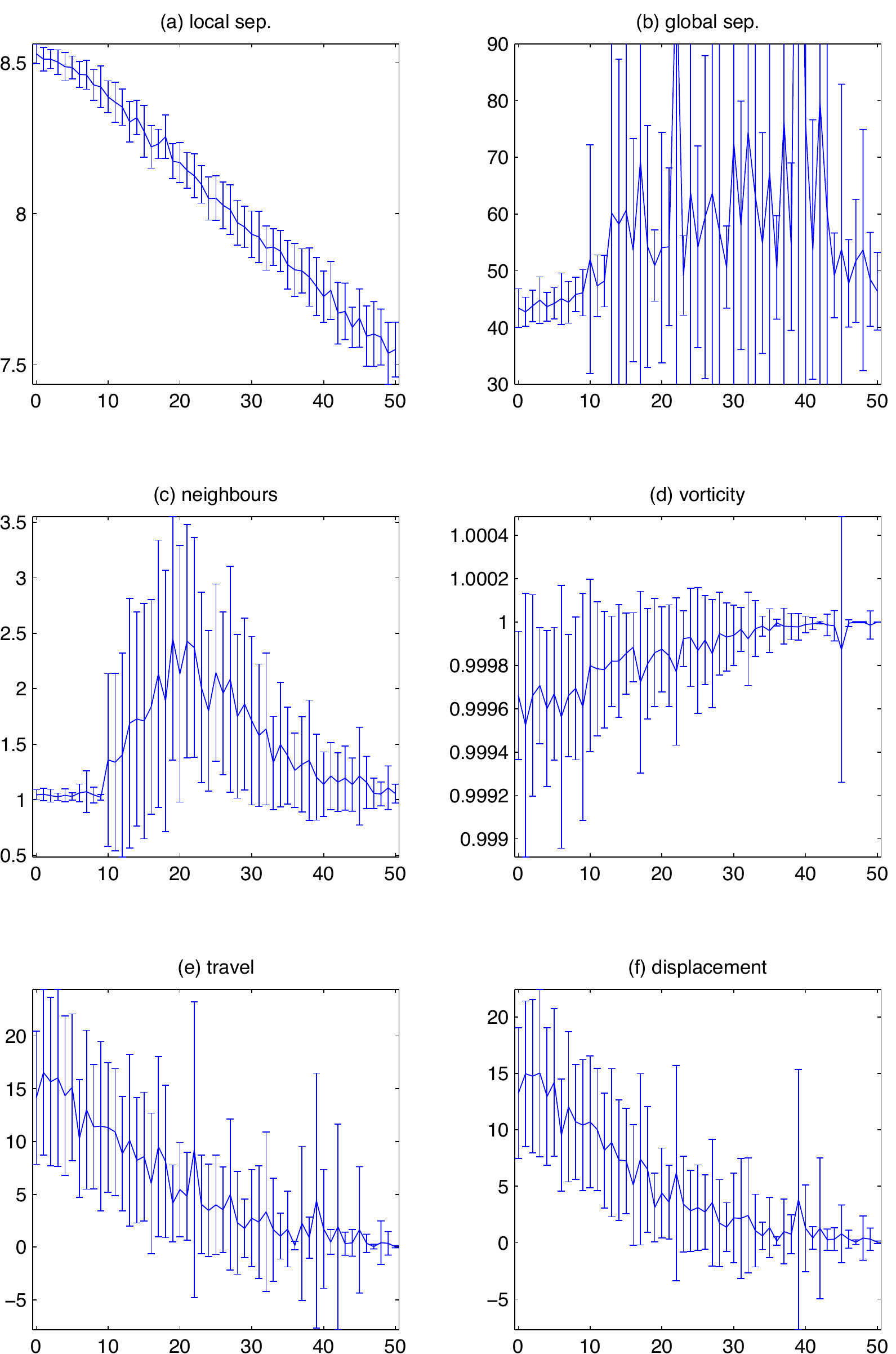}
\end{tabular}
\end{center}
\caption{{\bf Followers} For fixed $k$ and $N$ we now varying the number of followers $k$ in the swarm. Here we illustrate with results for $N=50$ and $k=6$. We set $r=0,1,2,\ldots 50$ and for each fixed $r$ simulate, from random initial conditions a swarm for $20000$ time steps, reporting dynamical measures over the lat $500$ time step (sampling every $10$-th). Mean and standard deviation of $30$ trials are shown.}
\label{followers}
\end{figure}

While we have identified a transition to instability which can occur for a range of $N$ and $k$, the stable configurations are both deterministic and rigid. In this respect they are, perhaps, not a good proxy for real animal behavior. Hence, it is natural to ask what would  need to be added to the model to introduce the natural variability of the real world. Stochastic perturbation is an obvious answer which we choose to ignore. Rather, we focus of whether introducing a small amount of aberrant (inhomogeneous) behavior is sufficient to make a previously stable configuration unstable. While the effect of leaders in swarm simulations has been widely studied (particularly from the perspective of control) we choose to introduce aberrant  ``followers''. In a model such as the one we are looking at here, it is fairly straightforward to see that a sentient leader will easily be able to direct the motion of the flock. Rather, we ask what happens if we introduce a small number of rigid neighbor links. That is, rather that the neighbors of $x^{(i)}$ being the particles $x^{(j)}$ such that $d_{ij}=\|x^{(i)}-x^{(j)}\|$ is least, we simply prescribe that particle $x^{(\ell(i))}$ is and always must be one of the neighbors of $x^{(i)}$.  The distance $d_{i\ell(i)}$ may not necessarily be smallest, particle $x^{(\ell(i))}$ is a ``neighbor'' of $x^{(i)}$ only for the purposes of computing the update rules (\ref{vh}-\ref{vv}). Moreover, the relationship is not necessarily reciprocated: particle $x^{(\ell(i))}$ has no compulsion to consider $x^{(i)}$ as a neighbor --- unless they actually are close. Finally, we do this for a fixed small number of neighbors $r\ll Nk$. If $r=k$ then each particle is a follower of exactly one other random particle, iff $r>k$ will a particle be a follower of more than one other particle.

Figure \ref{followers} depicts the effect of introducing these followers to our modeling scheme. Local separation shows a clear linear decrease with increasing $r$. Global separation, however, increases dramatically. Beyond about $r=12$ the swarm will fracture and form separate sub-groups. This behavior persists until $r>43$. Most dramatically, from Fig. \ref{followers}(c) we see a sudden transition at $r=10$ --- from a rigid solid state to a gaseous one. Increasing $r\rightarrow 50$ eventually reverses this trend. Finally, from Fig. \ref{followers} (d-f) we see that as the gaseous phase occurs the swarm ceases to rotate and overall movement is less.

We have shown that movement interaction rules adapted from nature can exhibit a wide range of interesting collective phenomena. The swarming behavior is notably coherent (forming a single flock) and exhibits a phase transition from a rigid crystalline state to a gaseous fluid-like one as a function of flock size and local neighborhood size. Nonetheless, the stable configuration is dominant and will re-emerge by further increasing the neighborhood size $k$. For a given stable configuration (fixed $N$ and $k$) we demonstrate that the introduction of a small number of followers leads to a threshold phenomena. Beyond a moderate value of $r$ ($N=50$ and $k=6$ we demonstrate that $r=10$) the swarm behavior again transitions to a gaseous phase. This persists until for very large $r$ the flock again becomes stable, possibly fracturing. Notably, the various dynamical behaviors we have observed here are deterministic. We have been able to offer a simple interaction model which mimics the variability found in nature without resorting to a probabilistic model. It would, perhaps, be interesting to extend this system to the case where the followers are not configured randomly but target specific individuals --- perhaps following a scale-free distribution.


%


\begin{thebibliography}{11}%
\makeatletter
\providecommand \@ifxundefined [1]{%
 \@ifx{#1\undefined}
}%
\providecommand \@ifnum [1]{%
 \ifnum #1\expandafter \@firstoftwo
 \else \expandafter \@secondoftwo
 \fi
}%
\providecommand \@ifx [1]{%
 \ifx #1\expandafter \@firstoftwo
 \else \expandafter \@secondoftwo
 \fi
}%
\providecommand \natexlab [1]{#1}%
\providecommand \enquote  [1]{``#1''}%
\providecommand \bibnamefont  [1]{#1}%
\providecommand \bibfnamefont [1]{#1}%
\providecommand \citenamefont [1]{#1}%
\providecommand \href@noop [0]{\@secondoftwo}%
\providecommand \href [0]{\begingroup \@sanitize@url \@href}%
\providecommand \@href[1]{\@@startlink{#1}\@@href}%
\providecommand \@@href[1]{\endgroup#1\@@endlink}%
\providecommand \@sanitize@url [0]{\catcode `\\12\catcode `\$12\catcode
  `\&12\catcode `\#12\catcode `\^12\catcode `\_12\catcode `\%12\relax}%
\providecommand \@@startlink[1]{}%
\providecommand \@@endlink[0]{}%
\providecommand \url  [0]{\begingroup\@sanitize@url \@url }%
\providecommand \@url [1]{\endgroup\@href {#1}{\urlprefix }}%
\providecommand \urlprefix  [0]{URL }%
\providecommand \Eprint [0]{\href }%
\providecommand \doibase [0]{http://dx.doi.org/}%
\providecommand \selectlanguage [0]{\@gobble}%
\providecommand \bibinfo  [0]{\@secondoftwo}%
\providecommand \bibfield  [0]{\@secondoftwo}%
\providecommand \translation [1]{[#1]}%
\providecommand \BibitemOpen [0]{}%
\providecommand \bibitemStop [0]{}%
\providecommand \bibitemNoStop [0]{.\EOS\space}%
\providecommand \EOS [0]{\spacefactor3000\relax}%
\providecommand \BibitemShut  [1]{\csname bibitem#1\endcsname}%
\let\auto@bib@innerbib\@empty
\bibitem [{\citenamefont {Sumpter}(2010)}]{dS10}%
  \BibitemOpen
  \bibfield  {author} {\bibinfo {author} {\bibfnamefont {D.~J.}\ \bibnamefont
  {Sumpter}},\ }\href@noop {} {\emph {\bibinfo {title} {Collective animal
  behavior}}}\ (\bibinfo  {publisher} {Princeton University Press},\ \bibinfo
  {year} {2010})\BibitemShut {NoStop}%
\bibitem [{\citenamefont {Reynolds}(1987)}]{cR87}%
  \BibitemOpen
  \bibfield  {author} {\bibinfo {author} {\bibfnamefont {C.~W.}\ \bibnamefont
  {Reynolds}},\ }in\ \href@noop {} {\emph {\bibinfo {booktitle} {Computer
  Graphics}}},\ Vol.~\bibinfo {volume} {21},\ \bibinfo {editor} {edited by\
  \bibinfo {editor} {\bibfnamefont {M.~C.}\ \bibnamefont {Stone}}}\ (\bibinfo
  {organization} {SIGGRAPH'87},\ \bibinfo {year} {1987})\ pp.\ \bibinfo {pages}
  {25--34}\BibitemShut {NoStop}%
\bibitem [{\citenamefont {Vicsek}\ \emph {et~al.}(1995)\citenamefont {Vicsek},
  \citenamefont {Czurk\'o}, \citenamefont {Ben-Jacob}, \citenamefont {Cohen},\
  and\ \citenamefont {Shochet}}]{tV95}%
  \BibitemOpen
  \bibfield  {author} {\bibinfo {author} {\bibfnamefont {T.}~\bibnamefont
  {Vicsek}}, \bibinfo {author} {\bibfnamefont {A.}~\bibnamefont {Czurk\'o}},
  \bibinfo {author} {\bibfnamefont {E.}~\bibnamefont {Ben-Jacob}}, \bibinfo
  {author} {\bibfnamefont {I.}~\bibnamefont {Cohen}}, \ and\ \bibinfo {author}
  {\bibfnamefont {O.}~\bibnamefont {Shochet}},\ }\href@noop {} {\bibfield
  {journal} {\bibinfo  {journal} {Physical Review Letters}\ }\textbf {\bibinfo
  {volume} {75}},\ \bibinfo {pages} {1226} (\bibinfo {year}
  {1995})}\BibitemShut {NoStop}%
\bibitem [{\citenamefont {Nagy}\ \emph {et~al.}(2010)\citenamefont {Nagy},
  \citenamefont {\'Akos}, \citenamefont {Biro},\ and\ \citenamefont
  {Vicsek}}]{mN10}%
  \BibitemOpen
  \bibfield  {author} {\bibinfo {author} {\bibfnamefont {M.}~\bibnamefont
  {Nagy}}, \bibinfo {author} {\bibfnamefont {Z.}~\bibnamefont {\'Akos}},
  \bibinfo {author} {\bibfnamefont {D.}~\bibnamefont {Biro}}, \ and\ \bibinfo
  {author} {\bibfnamefont {T.}~\bibnamefont {Vicsek}},\ }\href@noop {}
  {\bibfield  {journal} {\bibinfo  {journal} {Nature}\ }\textbf {\bibinfo
  {volume} {464}},\ \bibinfo {pages} {890} (\bibinfo {year}
  {2010})}\BibitemShut {NoStop}%
\bibitem [{\citenamefont {Dieck-Kattas}\ \emph {et~al.}(2012)\citenamefont
  {Dieck-Kattas}, \citenamefont {ke~Xu},\ and\ \citenamefont {Small}}]{chano1}%
  \BibitemOpen
  \bibfield  {author} {\bibinfo {author} {\bibfnamefont {G.}~\bibnamefont
  {Dieck-Kattas}}, \bibinfo {author} {\bibfnamefont {X.}~\bibnamefont {ke~Xu}},
  \ and\ \bibinfo {author} {\bibfnamefont {M.}~\bibnamefont {Small}},\
  }\href@noop {} {\bibfield  {journal} {\bibinfo  {journal} {PLoS Computational
  Biology}\ }\textbf {\bibinfo {volume} {8}},\ \bibinfo {pages} {e1002449}
  (\bibinfo {year} {2012})}\BibitemShut {NoStop}%
\bibitem [{\citenamefont {Xu}\ \emph {et~al.}(2012)\citenamefont {Xu},
  \citenamefont {Dieck-Kattas},\ and\ \citenamefont {Small}}]{xu5}%
  \BibitemOpen
  \bibfield  {author} {\bibinfo {author} {\bibfnamefont {X.-K.}\ \bibnamefont
  {Xu}}, \bibinfo {author} {\bibfnamefont {G.}~\bibnamefont {Dieck-Kattas}}, \
  and\ \bibinfo {author} {\bibfnamefont {M.}~\bibnamefont {Small}},\
  }\href@noop {} {\bibfield  {journal} {\bibinfo  {journal} {Physical Review
  E}\ }\textbf {\bibinfo {volume} {85}},\ \bibinfo {pages} {026120} (\bibinfo
  {year} {2012})}\BibitemShut {NoStop}%
\bibitem [{\citenamefont {Ward}\ \emph {et~al.}(2011)\citenamefont {Ward},
  \citenamefont {Herbert-Read}, \citenamefont {Sumpter},\ and\ \citenamefont
  {Krause}}]{aW11}%
  \BibitemOpen
  \bibfield  {author} {\bibinfo {author} {\bibfnamefont {A.~J.}\ \bibnamefont
  {Ward}}, \bibinfo {author} {\bibfnamefont {J.~E.}\ \bibnamefont
  {Herbert-Read}}, \bibinfo {author} {\bibfnamefont {D.~J.}\ \bibnamefont
  {Sumpter}}, \ and\ \bibinfo {author} {\bibfnamefont {J.}~\bibnamefont
  {Krause}},\ }\href@noop {} {\bibfield  {journal} {\bibinfo  {journal}
  {Proceedings of the National Academy of Sciences USA}\ }\textbf {\bibinfo
  {volume} {108}},\ \bibinfo {pages} {2312} (\bibinfo {year}
  {2011})}\BibitemShut {NoStop}%
\bibitem [{\citenamefont {Ballerini}\ \emph {et~al.}(2008)\citenamefont
  {Ballerini}, \citenamefont {Cabibbo}, \citenamefont {Candelier},
  \citenamefont {Cavagna}, \citenamefont {Cisbani}, \citenamefont {Giardina},
  \citenamefont {Orlando}, \citenamefont {Parisi}, \citenamefont {Procaccini},
  \citenamefont {Viale},\ and\ \citenamefont {Zdravkovic}}]{mB08}%
  \BibitemOpen
  \bibfield  {author} {\bibinfo {author} {\bibfnamefont {M.}~\bibnamefont
  {Ballerini}}, \bibinfo {author} {\bibfnamefont {N.}~\bibnamefont {Cabibbo}},
  \bibinfo {author} {\bibfnamefont {R.}~\bibnamefont {Candelier}}, \bibinfo
  {author} {\bibfnamefont {A.}~\bibnamefont {Cavagna}}, \bibinfo {author}
  {\bibfnamefont {E.}~\bibnamefont {Cisbani}}, \bibinfo {author} {\bibfnamefont
  {I.}~\bibnamefont {Giardina}}, \bibinfo {author} {\bibfnamefont
  {A.}~\bibnamefont {Orlando}}, \bibinfo {author} {\bibfnamefont
  {G.}~\bibnamefont {Parisi}}, \bibinfo {author} {\bibfnamefont
  {A.}~\bibnamefont {Procaccini}}, \bibinfo {author} {\bibfnamefont
  {M.}~\bibnamefont {Viale}}, \ and\ \bibinfo {author} {\bibfnamefont
  {V.}~\bibnamefont {Zdravkovic}},\ }\href@noop {} {\bibfield  {journal}
  {\bibinfo  {journal} {Animal Behaviour}\ }\textbf {\bibinfo {volume} {76}},\
  \bibinfo {pages} {201} (\bibinfo {year} {2008})}\BibitemShut {NoStop}%
\bibitem [{\citenamefont {Judd}\ and\ \citenamefont {Mees}(1995)}]{kJ95a}%
  \BibitemOpen
  \bibfield  {author} {\bibinfo {author} {\bibfnamefont {K.}~\bibnamefont
  {Judd}}\ and\ \bibinfo {author} {\bibfnamefont {A.}~\bibnamefont {Mees}},\
  }\href@noop {} {\bibfield  {journal} {\bibinfo  {journal} {Physica D}\
  }\textbf {\bibinfo {volume} {82}},\ \bibinfo {pages} {426} (\bibinfo {year}
  {1995})}\BibitemShut {NoStop}%
\bibitem [{\citenamefont {Small}\ and\ \citenamefont
  {Judd}(1998)}]{model_thing}%
  \BibitemOpen
  \bibfield  {author} {\bibinfo {author} {\bibfnamefont {M.}~\bibnamefont
  {Small}}\ and\ \bibinfo {author} {\bibfnamefont {K.}~\bibnamefont {Judd}},\
  }\href@noop {} {\bibfield  {journal} {\bibinfo  {journal} {Physica D}\
  }\textbf {\bibinfo {volume} {117}},\ \bibinfo {pages} {283} (\bibinfo {year}
  {1998})}\BibitemShut {NoStop}%
\bibitem [{\citenamefont {Small}\ and\ \citenamefont {Tse}(2002)}]{mdlnn}%
  \BibitemOpen
  \bibfield  {author} {\bibinfo {author} {\bibfnamefont {M.}~\bibnamefont
  {Small}}\ and\ \bibinfo {author} {\bibfnamefont {C.}~\bibnamefont {Tse}},\
  }\href@noop {} {\bibfield  {journal} {\bibinfo  {journal} {Physical Review
  E}\ }\textbf {\bibinfo {volume} {66}},\ \bibinfo {pages} {066701} (\bibinfo
  {year} {2002})}\BibitemShut {NoStop}%
\end{thebibliography}

\end{document}